\documentstyle{jfm}
\RequirePackage{epsfig}

\newcommand{\beq}{\begin{equation}}
\newcommand{\eeq}{\end{equation}}
\newcommand{\beqarr}{\begin{eqnarray}}
\newcommand{\eeqarr}{\end{eqnarray}}
\newcommand{\barr}{\begin{array}}
\newcommand{\earr}{\end{array}}
\newcommand{\bcent}{\begin{center}}
\newcommand{\ecent}{\end{center}}
\newcommand{\rf}[1]{(\ref{#1})}

\renewcommand{\vec}[1]{\mbox{\boldmath $#1$}}
\newcommand{\vechat}[1]{{\skew3\hat{\vec{#1}}}}
\newcommand{\mat}[1]{\mbox{\sls #1}}
\newcommand{\laplace}{\nabla^2}  
\newcommand{\laplacev}{\bnabla^2}
\newcommand{\laplacet}{\tilde{\nabla}^2}
\newcommand{\cross}{\wedge}
\newcommand{\grad}{\bnabla}
\newcommand{\dvgnce}{\bnabla \cdot}
\newcommand{\curl}{\bnabla \wedge}

\newcommand{\mod}[1]{\, |#1| \,}

\newcommand{\pd}[1]{\upartial_{#1}}
\newcommand{\dt}{{\Delta t}}
\newcommand{\bess}{{\mathcal{B}}}
\newcommand{\ex}{{\mathrm e}}
\newcommand{\im}{{\mathrm i}}

\newcommand{\on}{\mbox{{\em\space}on{\em\space}}}
\newcommand{\orr}{\mbox{{\em\space}or{\em\space}}}

\newcommand{\uc}{\tilde{\vec{u}}}	

\newcommand{\uct}{\tilde{u}_\theta}

\newcommand{\vel}{\vec{u}}		
\newcommand{\magn}{\vec{B}}
\newcommand{\nlin}{\vec{N}}
\newcommand{\tor}{{\mathcal{T}}}
\newcommand{\pol}{{\mathcal{P}}}
\newcommand{\prandtl}{\xi}		
\newcommand{\hartmann}{Q}
\newcommand{\eqnlabel}[1]{\eqno{(\theequation {\it #1})}}
\newcommand{\eqntext}[1]{$$ \resetline #1 $$}

\newenvironment{eqnalph}
   {\refstepcounter{equation} \let\\=\eqnalpharr $$}
   {$$ \resetline}

\newcommand{\resetline}{\vspace{-\baselineskip}\newline}
\def\eqnalpharr[#1]{$$\vspace{-10pt}\vspace{#1}\resetline$$}

\catcode`\œ=\active \gdefœ{\setbox0=\hbox{0}\hbox to\wd0{}}
\def\Rey{\mbox{\it Re}}   
\ifCUPmtlplainloaded
\else
  \def\upi{\pi} 
  \def\upartial{\partial} 
\fi
\ifCUPmtlplainloaded

\fi
\ifCUPmtlplainloaded
  \font\sls = mtssbi10 at 10.5pt  
  \font\bit = mtmib10 at 10.5pt \skewchar\bit ='177  
\else
  \font\sls = cmssi10  
  \font\bit = cmmib10 \skewchar\bit ='177  
\fi
\ifCUPmtlplainloaded
\else
  \font\tenbmi=cmmib10 at 10pt  \skewchar\tenbmi ='177
  \font\sevenbmi=cmmib10 at 7pt \skewchar\sevenbmi ='177
  \font\fivebmi=cmmib10 at 5pt  \skewchar\fivebmi ='177

  \newfam\bmifam
  \textfont\bmifam=\tenbmi
  \scriptfont\bmifam=\sevenbmi
  \scriptscriptfont\bmifam=\fivebmi
  
\fi
\newsavebox{\thalfbox}
\sbox{\thalfbox}{$\textstyle\frac{1}{2}$}

\newsavebox{\shalfbox}
\sbox{\shalfbox}{$\scriptstyle\frac{1}{2}$}

\newsavebox{\squartbox}
\sbox{\squartbox}{$\frac{1}{4}$} 

\newsavebox{\etbox}
\sbox{\etbox}{\boldmath$\eta$}

\newsavebox{\astrutbox}
\sbox{\astrutbox}{\rule[-5pt]{0pt}{20pt}}

\ifnfsstwo

\fi
\ifnfssone
  \newmathalphabet{\mathit}
    \addtoversion{normal}{\mathit}{cmr}{m}{it}
    \addtoversion{bold}{\mathit}{cmr}{bx}{it}

\fi
\ifoldfss

\fi

\mathchardef\varLambda="0103

\ifCUPmtlplainloaded
\else
\fi
\ifCUPmtlplainloaded
  \let\bcdot=\undefined
  \NewSymbolFont{bldsym}{mtbsy10}{'60}
  \NewMathSymbol{\bcdot}{2}{bldsym}{01}
\else
  \font\tenbms=cmbsy10          \skewchar\tenbms ='60
  \font\sevenbms=cmbsy10 at 7pt \skewchar\sevenbms ='60
  \font\fivebms=cmbsy10 at 5pt  \skewchar\fivebms ='60

  \newfam\bmsfam
  \textfont\bmsfam=\tenbms
  \scriptfont\bmsfam=\sevenbms
  \scriptscriptfont\bmsfam=\fivebms

  \edef\bsy{\hexnumber\bmsfam}
  \mathchardef\bnabla="0\bsy72
  \mathchardef\bcdotsymbol="0\bsy01
  \def\bcdot{\,\bcdotsymbol\,}
\fi
\def\eg{{e.g.\ }}

\def\etal{\mbox{\it et al.\ }}

\title[Hydromagnetic Taylor--Couette flow]
{Hydromagnetic Taylor--Couette flow: \\
numerical formulation and comparison with experiment}

\author[A. P. Willis and C. F. Barenghi]%
{
   A.\ns P.\ns W\ls I\ls L\ls L\ls I\ls S\ns 
   \and 
   C.\ns F.\ns B\ls A\ls R\ls E\ls N\ls G\ls H\ls I\ls
}

\affiliation
{
   Department of Mathematics, University of Newcastle,\\
   Newcastle NE1 7RU, England
}

\pubyear{2001}
\volume{}
\pagerange{}
\date{July 2001}
\begin{document}

\maketitle

\begin{abstract}
Taylor--Couette flow in the presence of a magnetic field 
is a problem belonging to classical hydromagnetics and deserves 
to be more widely studied than it has been to date.  
In the nonlinear regime the literature is scarce.
We develop a formulation 
suitable for solution of the full three dimensional nonlinear 
hydromagnetic equations in cylindrical geometry, which is motived
by the formulation for the magnetic field. 
It is suitable for study at finite Prandtl numbers 
and in the small Prandtl number limit, relevant to laboratory liquid metals.
The method is used to determine the onset of axisymmetric Taylor vortices, 
and finite amplitude solutions.  
Our results compare well with existing 
linear and nonlinear hydrodynamic calculations and with hydromagnetic
experiments.
\end{abstract}

\section{Introduction}

The motion of an incompressible viscous fluid between concentric rotating
cylinders is one of the most important
problems of fluid dynamics and is much studied as a benchmark to
investigate issue of instability and nonlinear behaviour. 
\cite{taylor23} found that if the rotation of the inner 
cylinder is greater than some critical value then circular--Couette 
flow (CCF) becomes unstable to axisymmetric perturbations.  
A secondary flow appears which has axial and radial motion in the 
form of pairs of toroidal vortices, now known as the Taylor--vortex 
flow (TVF).  
If the inner cylinder is driven further then this flow becomes unstable 
to non-axisymmetric perturbations.  Azimuthal waves appear in the 
Taylor--vortices and the whole pattern rotates at some wavespeed
(wavy modes).

In his
landmark 1961 book on stability theory, Chandrasekhar
devoted equal attention to the hydrodynamic and the hydromagnetic
Couette problems; the latter is the case in which the fluid is a conducting
liquid (\eg mercury, liquid gallium, liquid sodium) and a magnetic
field is applied externally. Despite this early interest in the
hydromagnetic Couette problem, which included experiments performed by 
\cite{donnelly62} and by \cite{donnelly63}, 
most of the activity of the following years was devoted
to the hydrodynamic case. Among the few studies of the effects of the
magnetic field it is worth remembering the works by
\cite{velikhov59}, \cite{kurzweg63},
\cite{roberts64}, who extended Chandrasekhar's theory to non-axisymmetric
bifurcations from circular--Couette flow, \cite{chang67} and
\cite{hunt71} at finite aspect ratio.
Later \cite{tabeling81},
using a method similar to Davey's (1962) amplitude expansion,
calculated effective viscosity of axisymmetric flow in the Taylor vortex 
flow regime; he compared against Donnelly's (1962) experiments which 
indicate that the onset of wavy vortices is significantly inhibited by 
the magnetic field. \cite{nagata96} has more recently investigated nonlinear 
solutions in the planar geometry, and \cite{hollerbach00a} shows 
Taylor cells in spherical geometry.

The aim of this paper is to investigate effects induced on Couette
flow by an externally applied magnetic field. This paper is
meant to be the first of
a series and is dedicated to the development of a suitable formulation for
solving numerically the governing nonlinear three dimensional 
magnetohydrodynamic (MHD) equations in the cylindrical Couette geometry.  
The numerical method
which we propose can be used for any value of radius ratio, is suitable
for time stepping, has good stability features, is relatively easy to
program and is more accurate than existing methods.

Our work is also motivated by the
renewed interest in MHD flows in confined geometries which
arises from current and planned experiments to produce dynamo action in
the laboratory 
(Gailitis \etal 2001; 
 Stieglitz \& Muller 2001).
It must be stressed that our work does not apply directly to the dynamo
problem for two reasons.
Firstly the cylindrical Couette configuration is a possible geometry
for these studies, but it has not been used in experiments yet, and it may
prove not to be not the most efficient one
(Laure, Chossat \& Daviaud, 2000).  
Secondly, our investigation refers primarily to bifurcations at relatively
small Reynolds numbers, while the dynamo experiments
require rather large Reynolds numbers due to the small
magnetic Prandtl number of liquid metals.
Despite these two limitations, our work is related to the dynamo problem
because it is important to have precise results at small Reynolds 
number MHD flows in order to develop and test modern acoustic flow 
visualisation techniques, \cite{kikura99}, 
which offer the best chance to
detect flow patterns in MHD dynamos. The lack of flow visualisation 
has clearly held back progress in the hydromagnetic Couette problem 
compared to the hydrodynamic case.

\section{Equations}
\label{sect:eqns}

The equations governing incompressible hydromagnetic flow are
\begin{eqnalph}
   \label{eq:gov}
   \pd{t} \vel + (\vel \cdot \grad) \vel = 
   - \frac{1}{\rho} \grad p + \nu \laplacev \vel 
   + \frac{1}{\rho\mu_0} (\curl \magn) \cross \magn,
   \qquad
   \dvgnce \vel = 0,
   \eqnlabel{a,b}
   \\[0pt]
   \pd{t} \magn =
   \lambda \laplacev \magn + \curl (\vel \cross \magn),
   \qquad
   \dvgnce \magn = 0,
   \eqnlabel{c,d}
\end{eqnalph}
where $\vel$ is the fluid's velocity, $\magn$ the magnetic field,
$p$ the pressure, $\rho$ the density,
$\nu$ the kinematic viscosity, $\lambda$  the
magnetic diffusivity and $\mu_0$ the magnetic permeability.
Hereafter we assume that $\rho$, $\nu$, $\lambda$ and $\mu_0$ are constant.
The fluid is contained between 
two concentric cylinders of inner radius $R_1$ and outer
radius $R_2$. The inner and outer cylinder rotate at
constant angular velocities $\Omega_1$ and $\Omega_2$ respectively.
A magnetic field $\magn_0=\mu_0H\vechat{z}$
is applied externally in the axial direction.
We make the usual simplifying assumption that 
the cylinders have infinite length and use
cylindrical coordinates $(r,\theta,z)$.

Throughout the rest of this work we will make the variables
dimensionless using the following scales:
\[
   \barr{llll}
      \delta = R_2 - R_1,	&  \mbox{length (gap width)};		&
      \delta^2 / \nu,		&  \mbox{time (viscous diffusion)};	\\
      \nu / \delta,		&  \mbox{velocity};                     &
      \mu_0H                 	&  \mbox{magnetic field}.
   \earr 
\]
We introduce the following dimensionless parameters: radius ratio ($\eta$),
Reynolds numbers ($\Rey_1$ and $\Rey_2$), Hartmann number ($Q$) and magnetic
Prandtl number ($\xi$) defined as
   \beq
      \eta = R_1/R_2;  \quad 
      \Rey_i = \frac{R_i \Omega_i \delta}{\nu},  \quad
      i=1,2; \quad
      \hartmann = \frac{\mu_0^2 H^2 \sigma \delta^2}{\rho \nu};  \qquad
      \prandtl = \frac{\nu}{\lambda} .
   \eeq
The dimensionless forms of (\ref{eq:gov}{\it a,c}) are then
\begin{eqnalph}
   \label{eq:non-d_gov}
   \pd{t} \vel + (\vel \cdot \grad) \vel =
   - \grad p + \laplacev \vel 
   + \frac{\hartmann}{\prandtl} (\grad \cross \magn) \cross \magn,
   \eqnlabel{a}
   \\[0pt]
   \pd{t} \magn  = 
   \frac{1}{\prandtl} \laplacev \magn 
   + \grad \cross (\vel \cross \magn) .
   \eqnlabel{b}
\end{eqnalph}
A steady-state solution of the governing equations 
is circular--Couette flow, $\uc=(0,\uct,0)$ where $\uct$ is
$ \uct = Ar + B/r $.
The constants $A$ and $B$ are determined by the no-slip boundary 
conditions.  We set $\vel = \uc+\vel'$ and $p=\tilde{p}+p'$.
The deviation, $\vel'$, then satisfies the homogeneous Dirichlet
boundary condition,
      $\vel'=\vec{0}$ at $R_1, R_2$.
Subtracting the Navier--Stokes equation for $\uc$ from 
(\ref{eq:non-d_gov}{\it a}), the evolution of $\vel'$ is now described by
\begin{eqnalph}
   \label{eq:MHDeqns}
   (\pd{t} - \laplacev) \vel'  = 
   \nlin - \grad p',
   \qquad
   \dvgnce \vel' = 0,
   \eqnlabel{a,b}
   \\[0pt]
   ( \pd{t} - \frac{1}{\prandtl} \laplacev ) \magn  =  \nlin_B,
   \qquad
   \dvgnce \magn = 0,
   \eqnlabel{c,d}
\eqntext{with nonlinear terms,}
   \nlin =
   \frac{\hartmann}{\prandtl} (\curl \magn)\cross \magn
   - (\vel\cdot\grad) \vel' - (\vel'\cdot\grad) \uc,
   \qquad
   \nlin_B = \curl (\vel \cross \magn) .
   \eqnlabel{e,f}
\end{eqnalph}

The magnetic Prandtl number $\xi$ is very small in liquid metals
available in the laboratory, so we set
     $ \magn = \magn_0 + \prandtl \vec{b} $.
In the limit $\prandtl \to 0$ the governing equations become
\begin{eqnalph}
   \label{eq:gov_pr0}
   (\pd{t} - \laplacev) \vel' = \nlin - \grad p',
   \qquad
   \dvgnce \vel'  =  0,
   \eqnlabel{a,b}
   \\[-5pt]
   \laplacev \vec{b}  =  \nlin_B,
   \qquad
   \dvgnce \vec{b} = 0,
   \eqnlabel{c,d}
\eqntext{where,}
   \nlin  = 
   \hartmann (\curl \vec{b})\cross \magn_0
   - (\vel\cdot\grad) \vel' - (\vel'\cdot\grad) \uc,
   \qquad
   \nlin_B  = 
   - \curl (\vel \cross \magn_0) .
   \eqnlabel{e,f}
\end{eqnalph} 
Note that these equations are descriptive rather than predictive for
   $\vec{b}$.

\section{Boundary conditions}
   \label{sect:BCs}

The governing equations \rf{eq:MHDeqns} represent a tenth order 
system in $r$ and we therefore require ten boundary conditions.
The first six are simply the no-slip condition, $u_r=u_\theta=u_z=0$
applied at the boundaries $r=R_1$ and $r=R_2$. The boundary conditions
for the magnetic field depend on the nature of the container, as
discussed by \cite{roberts64}, who determined  conditions for
arbitrary values of electrical conductivity.  The experiments by
\cite{donnelly62} used mercury with Perspex and stainless--steel
containers.  Only a small difference was found between the results
obtained using different containers.  Hereafter we consider
only the simple case of insulating boundaries.  

Ampere's law says that,
$ \vec{J} = \prandtl^{-1} \curl \magn = 0$, when $r<R_1$ or $r>R_2$,
as the current within an insulator must be zero.
It follows that the magnetic field is irrotational and can be expressed
in terms of a potential, $\psi$, in the following way:
   \beq
      \magn = - \grad \psi,
      \qquad
      - \dvgnce \magn = \laplace \psi = 0.
   \eeq
This equation can be solved for $\psi$ by separation of variables,
     $ \psi(r,\theta,z) = R(r) \, \Theta(\theta) \, Z(z) $.
In our periodic coordinates we obtain
   \beq
      \frac{\Theta''(\theta)}{\Theta(\theta)} = - m^2,
      \quad
      \frac{Z''(z)}{Z(z)} = - \alpha^2,
   \eeq
where $m$ is integer.
The equation for $R(r)$ satisfies the modified Bessel equation,
   \beq
      \frac{1}{r}\, R'(r)+R''(r) - 
      \left(
         \alpha^2 + \frac{m^2}{r^2}
      \right) R(r) = 0 .
   \eeq
The boundary conditions for $R(r)$ then depend on the type of solution.
  
If $\psi$ is independent of $\theta$ and $z$ ($m=\alpha=0$) then
$\psi$ must be constant and so $\magn=\vec{0}$.  But this means that
we have three conditions at each boundary, and we only need two.
However, the divergence-free condition implies that a solution
which is independent of $\theta$ and $z$ must have no
radial component.  It is therefore sufficient to take
   \beq
      \label{eq:mag_bcs_k=m=0}
      B_\theta = B_z = 0 .
   \eeq
   
If $\psi$ is independent of $z$ but depends on $\theta$ 
($\alpha=0, m\ne 0$) then $R(r)=r^{\pm m}$.  Recalling that 
$\magn=\grad\psi$, we have
   \beq
      \label{eq:mag_bcs_k=0}
      \pd{\theta} B_r = \pm m B_\theta,
      \quad
      B_z = 0 .
   \eeq
   
If $\psi$ is $z$ dependent ($\alpha\ne 0$) then $R(r) = \bess_m(r)$
where $\bess_m(r)$ denotes either of the modified Bessel functions
$I_m(\alpha r), K_m(\alpha r)$.  We obtain
   \beq
      \label{eq:mag_bcs}
      \pd{z}B_r = \frac{\pd{r}\bess_m}{\bess_m} \, B_z ,
      \quad
      \frac{1}{r} \, \pd{\theta} B_z = \pd{z} B_\theta .
   \eeq
In the outer region $r>R_2$ the field tends to zero, $\magn \to 0$ as
$r \to \infty$, and in the inner region $r<R_1$ $\magn$ must remain finite,
which implies that we take
   \beq
      \label{eq:Rfns}
      R(r) = 
      \left\{
         \barr{lll}
            I_m(\alpha r) \\ K_m(\alpha r)
         \earr
      \right.
      \orr
      \barr{l}
         r^{+m} \\ r^{-m}
      \earr
      \on
      \barr{l}
         r\le R_1 \\ r \ge R_2
      \earr
   \eeq
The two relations \rf{eq:mag_bcs_k=m=0}, \rf{eq:mag_bcs_k=0}, 
or \rf{eq:mag_bcs} applied at the points $r=R_1$ and $r=R_2$,
given the appropriate function from \rf{eq:Rfns},
are equivalent to Roberts' insulating boundary conditions.   
In this way we have the remaining four boundary conditions which
are required.

\section{Formulation and solution}

   The difficulty with primitive-variable formulations is how to ensure
   a divergence-free field.  A possible solution is to combine time 
   splitting and pressure projection.  The
   divergence of the momentum equation gives a Poisson equation
   for the pressure which is used to project the velocity into the space 
   of solenoidal functions.  
   No pressure term occurs naturally in the induction equation,
   and if not removed divergence can build up in the solution for the 
   magnetic field, especially at larger magnetic Prandtl numbers.  However,
   an arbitrary projection function could be added in order for the 
   divergence to scale with the timestep.  
   \cite{marcus84} used an influence matrix method in order
   to implement the correct boundary conditions for the pressure
   (Rempfer, 2002).
   This technique leads to a divergence that is zero to machine accuracy.
   In both methods an adjustment to the field is made at each 
   timestep.  However, for the hydromagnetic case there is no timestep 
   in the small Prandtl number limit; 
   without potentials it is difficult to invert the Poisson equation for 
   the magnetic field whilst simultaneously ensuring it is divergence-free.
   We propose a formulation able to cope with both finite Prandtl numbers
   and the small Prandtl number limit without significant adjustments.
  
   Popular in MHD is the toroidal-poloidal potentials form
   where variables are decomposed as
   $\vec{A} = \curl (\psi\vec{e}) + \curl \curl (\phi\vec{e})$
   where $\vec{e}$ is a vector constant.  
   To eliminate the pressure in the momentum equation it is commonplace to 
   take the $e$-components of the first and second curls
   as the governing equations for the velocity. 
   For the magnetic field it is sufficient to take the $e$-components
   of the induction equation and its first curl.
 
   In spherical geometry one assumes $\vec{e}=\vec{r}_s$, 
   the spherical radius.
   Taking $r_s$-components of successive curls, the Naiver--Stokes and 
   induction equations separate into one equation for each of the 
   potentials, \cite{hollerbach00b}.  
   Unfortunately complications can arise if this method is used in 
   cylindrical geometry.  The choice
   $\vec{e}=\vechat{z}$ leads to separate equations for each of the 
   potentials but raises the order of the equations in $r$. 
   However, \cite{marques90} has derived the extra boundary conditions 
   required for the hydrodynamic problem.

   Taking the second curl leads to an operator acting on one of
   the potentials in the form of double Laplacians.  
   In spherical geometry \cite{tilgner97} successfully implemented a 
   second order code using this formulation with stress-free boundaries.
   \cite{hollerbach00b} also used this formulation with no-slip boundary 
   conditions but found it to be unstable, even for very small 
   timesteps, unless an implicit first order time discretisation was used.
   In the cylindrical geometry \cite{rudiger00} used the formulation of 
   \cite{marques90}, but similarly used a first order method to avoid
   numerical difficulties.
   It is not at all clear that higher derivatives necessarily entail
   numerical instability in hydrodynamical solvers.  
   However, for reasons that become apparent 
   when the velocity field is discussed (\S\ref{sect:velfield}),
   the second curl is avoided.

   Instead we are motivated by a parallelism with the magnetic 
   field and take only the first curl.  
   As the pressure has not been eliminated, we also take the
   divergence of the momentum equation.  A single-curl formulation was
   proposed by \cite{glatzmaier84} in the spherical geometry and has
   proved very successful.

   We have made the choice $\vec{e}=\vec{r}$ where $\vec{r}$ is the 
   cylindrical-polar radius.  This choice gives equations which couple 
   the potentials, but this is no particular problem; with primitive
   variables $r$ and $\theta$ components are coupled.
   Fortunately, although we still
   raise the order of the equations, the extra derivatives appear in the 
   periodic coordinates and so no extra boundary conditions are required.
   To ensure capture of all possible solutions 
   extra terms along $\vechat{\theta}$ and $\vechat{z}$ are added,
   in order to accommodate solutions that are independent of both
   $\theta$ and $z$.  Full expansion of variables has the form
   \beq
      \label{eq:pot_expansion}
      \vec{A} = \psi_0 \, \vechat{\theta} + \phi_0 \, \vechat{z}
      + \curl (\psi\vec{r}) + \curl \curl (\phi\vec{r}),
   \eeq
   where $\psi(r,t,z)$, $\phi(r,t,z)$ and 
   $\psi_0(r)$, $\phi_0(r)$ contain the periodic and non-periodic parts
   respectively.
   We discuss first the formulation for the magnetic field, as it 
   motivates the method for the velocity.
%

\subsection{\bf The magnetic field}

   The magnetic field is expanded as
   \beq
      \magn = \tor_0 \, \vechat{\theta} + \pol_0 \, \vechat{z}
      + \curl (\tor\vec{r}) + \curl \curl (\pol\vec{r}),
   \eeq
   and substituted into the induction equation, 
   (\ref{eq:MHDeqns}{\it d}).  
   For the non-periodic potentials $\tor_0,\pol_0$ the governing 
   equations are obtained
   from the $\theta,z$ components
   \begin{eqnalph}
      (\pd{t} - \frac{1}{\prandtl}(\laplace-\frac{1}{r})) \tor_0 
      = \vechat{\theta} \cdot \nlin_B ,
      \eqnlabel{a}
      \\[0pt]
         (\pd{t} - \frac{1}{\prandtl}\laplace) \pol_0
         = \vechat{z} \cdot \nlin_B .
      \eqnlabel{b}
   \end{eqnalph}
   with boundary conditions at $R_1,R_2$,
   \beq
      \tor_0 = 0, \quad \pol_0 = 0.
   \eeq
   The periodic potentials $\tor,\pol$ are assumed to be of form
   $\ex^{\im(\alpha z+m\theta)}$.  In order to match the boundary
   conditions a spectral expansion will be required.
   There is no pressure term to eliminate here, so 
   we take the $r$-components of 
   the induction equation and its first curl
   \begin{eqnalph}
      \label{eq:mag_gov_eqns}
      \frac{2}{\prandtl r^2} \, \pd{\theta z} \tor 
      - \laplace_c (\pd{t} - \frac{1}{\prandtl}\laplacet) \pol
      - \frac{2}{\prandtl r^3} \, \pd{r\theta\theta} \pol
      = \frac{1}{r^2} \, \vec{r} \cdot \nlin_B ,
      \eqnlabel{a}
      \\[0pt]
      - \laplace_c (\pd{t} - \frac{1}{\prandtl}\laplacet) \tor
      - \frac{2}{\prandtl r^3} \, \pd{r\theta\theta} \tor
      + \frac{2}{r^2} (\pd{t}-\frac{2}{\prandtl}\laplace) 
      \pd{\theta z}\pol
      = \frac{1}{r^2} \, \vec{r} \cdot \curl \nlin_B ,
      \eqnlabel{b}
   \end{eqnalph}
   where
   \beq
      \laplacet   =  \laplace + \frac{2}{r} \, \pd{r} ,
      \qquad
      \laplace_c  =  \frac{1}{r^2}\,\pd{\theta\theta} + \pd{zz} .
   \eeq
   with boundary conditions at $R_1,R_2$,
   \begin{eqnalph}
      \label{eq:mag_pot_bcs}
      \alpha = 0:
      \qquad
      \pd{\theta} \tor = 0 ,
      \quad
      \left(
         \laplace_c \pm \frac{m}{r} \, \pd{r}
      \right)
      \pd{\theta} \pol = 0 ;
      \eqnlabel{a}
   \\[0pt]
      \alpha \ne 0:
      \qquad
      \left.
      \barr{l}
      {\displaystyle
         \laplace_c \tor - \frac{2}{r^2} \, \pd{\theta z}\pol = 0,
      }\\
      {\displaystyle
         \frac{1}{r}\,\pd{\theta}\tor -
         \left(
            \frac{\bess_m}{\pd{r}\bess_m}\, \laplace_c + 
            \frac{2}{r} + \pd{r}
         \right)
         \pd{z} \pol = 0
      } .
      \earr
      \right\}
      \eqnlabel{b}
   \end{eqnalph}

   This formulation is suitable for the small Prandtl number limit
   \rf{eq:gov_pr0}, relevant to laboratory liquid metals.
   We expand $\vec{b}$ by the same potentials and make the replacements,
   \[
      \pd{t} \to 0, 
      \qquad
      - \,\frac{1}{\prandtl} \to 1,
   \]
   throughout.  The field $\vec{b}$ satisfies the same boundary 
   conditions as $\magn$.

   Having settled on a formulation of the equations, we now discuss 
   the numerical method.

\subsection{\bf Numerical method}

   It is customary to adjust the radial range into the unit interval,
   so a new radial variable $x$ is defined as
   \beq
      r = R_1 + x, \qquad
      R_1 = \eta / (1-\eta), \qquad
      x \in [0,1].
   \eeq
   If a field is expected to have $m_1$-fold rotational symmetry, such as
   the case of wavy modes, then variables are expanded as
   \beq
      A(x,\theta,z,t) 
      = \sum_{n=0}^N \, \sum_{\mod{k}<K} \, \sum_{\mod{m}<M} \,
      A_{nkm}(t) \, T_n^*(x) \, \ex^{\im(\alpha kz + m_1 m\theta)}
   \eeq
   on the domain $[0,1]\times[0,2\upi/m_1]\times[0,2\upi/\alpha]$ 
   where $T_n^*(x)$ is the $n^{th}$ shifted Chebyshev polynomial.
   Variables are collocated on the $N+1$ extrema of $T_N(x)$.
   This arrangement of points is well suited to our problem with the 
   points concentrated near the boundaries.

   As the velocity and magnetic field are coupled by the
   nonlinear terms it makes sense to treat them explicitly.  
   Nonlinear terms are evaluated pseudospectrally, where necessary.
   Large terms in the zero-modes, like circular--Couette flow and the
   imposed magnetic field, can be extracted and calculated exactly.
   Let $q$ indicate the time discretisation $t_q=q\,\dt$ with $q=0,1,2,...$.
   We choose to use second order Adams--Bashforth to 
   estimate $\nlin_B$ at the intermediate time $q+\frac{1}{2}$.

   The linear terms are easier to evaluate and are timestepped using the
   implicit Crank--Nicolson method.
   Substituting the spectral expansion in the governing equations 
   \rf{eq:mag_gov_eqns} and the boundary conditions \rf{eq:mag_pot_bcs}, 
   after collocation the problem for each $k,m$ mode becomes
   \beq
      \mat{X} \left[\barr{l}\tor\\\pol\earr\right]^{q+1} 
      = \mat{Y} \left[\barr{l}\tor\\ \pol\earr\right]^q + 
      \left[\barr{l}N_1\\N_2\earr\right]^{q+\frac{1}{2}} ,
   \eeq
   where $\mat{X}$ and $\mat{Y}$ are matrices.  
   The vector $[\tor,\pol]^q$ contains the spectral coefficients 
   of the potentials at time $t_q$. 
   The nonlinear terms have been evaluated on the collocation points.
   Despite the Fourier expansions, the matrices $\mat{X},\mat{Y}$ are 
   \emph{real} and are calculated by the same routine, as they differ 
   only by a scalar constant, namely $\pm \dt$.  

   With this formulation modifications for the small Prandtl number 
   limit \rf{eq:gov_pr0} are relatively minor.  
   The magnetic field is now completely defined by the
   velocity at some particular time $t_q$, 
   For each $k,m$ mode the problem in matrix-vector form is now simply
   \beq
      \mat{X} \left[\barr{l}\tor\\\pol\earr\right]^q
      =
      \left[\barr{l}N_1\\N_2\earr\right]^q .
   \eeq
   The matrices $\mat{X}$ are calculated by the same routines above,
   as, again they only differ by scalar constants.

\subsection{\bf The velocity field}
   \label{sect:velfield}
 
   Unless there is an externally imposed velocity field, there is
   no non-periodic pressure to be eliminated.  The 
   non-periodic part of the velocity is then treated in the same manner 
   as the magnetic field, i.e.
   \begin{eqnalph}
      (\pd{t} - (\laplace-\frac{1}{r})) \psi_0 
      =  \vechat{\theta} \cdot \nlin ,
      \eqnlabel{a}
      \\[5pt]
      (\pd{t} - \laplace) \phi_0
      =  \vechat{z} \cdot \nlin ,
      \eqnlabel{b}
   \end{eqnalph}
   with boundary conditions
   \beq
      \psi_0 = \phi_0 = 0 ,
   \eeq
   on $R_1,R_2$.

   For the periodic part we
   follow the procedure applied to the magnetic 
   field and take only the first curl.  
   As the pressure has not been eliminated, we also take the
   divergence of the momentum equation.  
   The equations obtained are
   \begin{eqnalph}
      \label{eq:gov_vel}
      \frac{2}{r^2} \, \pd{\theta z} \psi
      - \laplace_c (\pd{t} - \laplacet) \phi
      - \frac{2}{r^3} \, \pd{r\theta\theta} \phi
      = \frac{1}{r^2} \, \vec{r} \cdot (\nlin - \grad p),
      \eqnlabel{a}
      \\[0pt]
      - \laplace_c (\pd{t} - \laplacet) \psi
      - \frac{2}{r^3} \, \pd{r\theta\theta} \psi
      + \frac{2}{r^2} (\pd{t}-2\laplace) \pd{\theta z}\phi
      = \frac{1}{r^2} \, \vec{r} \cdot \curl \nlin ,
      \eqnlabel{b}
      \\[0pt]
      \laplace p = \dvgnce \nlin .
      \eqnlabel{c}
   \end{eqnalph}
   These equations may look compicated enough, but taking the 
   second curl they are much worse!
   However, they simplify considerably for the axisymmetric problem.
   Note also the simplification that the linear differential operators
   on the left-hand sides of 
   (\ref{eq:mag_gov_eqns}{\it a,b}) and (\ref{eq:gov_vel}{\it a,b})
   are the same with $\xi\rightarrow 1$.   
   Fourth order derivatives have been avoided, otherwise
   as $|{\mathrm d}^p \,T_n^*(x) / {\mathrm d}x^p| = O(n^{2p})$ matrices 
   can become difficult to invert accurately with larger truncations.  
   
   Every governing equation is only second order in $r$, and therefore
   all equations have the same number of associated boundary conditions.
   This permits us to take the same radial truncation $N$ for all 
   variables, so all matrices are likewise of the same size.  
   This simplifies the actual implementation enormously!  
   In fact, we
   timestep the governing equations (\ref{eq:gov_vel}{\it a,b}), 
   the same way as the magnetic field,
   \beq
      \label{eq:gov_vel_mat}
      \mat{X} \left[\barr{l}\psi\\\phi\earr\right]^* 
      = \mat{Y} \left[\barr{l}\psi\\ \phi\earr\right]^q + 
      \left[\barr{l}N_1\\N_2\earr\right]^{q+\frac{1}{2}} ,
   \eeq
   and the matrices $\mat{X}^{-1}, \mat{X}^{-1}\mat{Y}$ may be 
   precomputed by the routine for the magnetic field as the equations 
   are the same, but for scalar constants.  
   Thus, linear terms for the velocity are also timestepped using 
   Crank--Nicolson.  Using this method, \cite{marcus84} found that a 
   numerical neutrally stable oscillation can occur at large wavenumbers.
   Fortunately this does not present a problem here, as the presence 
   of the magnetic field tends to reduce the natural wavenumber.

   Together, the evolution equations (\ref{eq:gov_vel}{\it a,b}) are 
   only fourth order in $r$ for the potentials, 
   but there are six boundary conditions for the velocity.  
   They are timestepped with boundary conditions $u_\theta=u_z=0$, or
   \begin{eqnalph}
      \label{eq:vel_BCs}
      r \pd{z} \psi + \pd{r\theta} \phi = 0,
      \quad
      - \pd{\theta}\psi + (2+r\pd{r})\,\pd{z} \phi = 0.
      \eqnlabel{a,b}
   \eqntext{
      The pressure--Poisson equation (\ref{eq:gov_vel}{\it c})
      is inverted with the boundary condition
      $u_r=0$, or,
   }
      \label{eq:press_bc}
      - r \laplace_c \phi = 0,
      \eqnlabel{c}
   \end{eqnalph}
   essentially the no-penetration condition.  

   Inversion of (\ref{eq:gov_vel}{\it c}) requires a 
   boundary condition, 
   indirectly determined by
   (\ref{eq:press_bc}{\it c}),
   in terms of $p$.  
   The adjustment for pressure is
   \beq
      \label{eq:p_adj}
      \left[\barr{l}\psi\\\phi\earr\right]^{q+1} 
      = \left[\barr{l}\psi\\\phi\earr\right]^* - 
      \mat{X}^{\,-1} \left[\barr{c}\frac{1}{r}\,\pd{r}p\\0\earr\right].
   \eeq
   In order to separate $\phi^{q+1}$ from $\psi^{q+1}$ we must work with
   $\mat{X}^{\,-1}$ rather than $\mat{X}$.
   For each $k,m$ mode $\laplace_c$ is just a scalar.
   Imposing $-r\laplace_c \phi^{q+1}=0$ the
   boundary condition becomes,
   \beq
      \label{eq:numerical_pBC}
      {\skew3\hat{\mat{X}}}^{\,-1}
      \left(
         \frac{1}{r} \, \pd{r}p
      \right)
      = \phi^* ,
   \eeq
   where ${\skew3\hat{\mat{X}}}^{\,-1}$ is the lower left quadrant of
   $\mat{X}^{\,-1}$.  Condition \rf{eq:numerical_pBC} is implemented
   in the usual manner by multiplying on the left with the coefficients 
   $T_n^*(x)$ at the boundaries.

\section{Results}


\subsection{\bf Linear stability}

   The linear part of the code is shared between the velocity and
   magnetic fields.  Appropriate 
   tests are determining the critical Reynolds number, $\Rey_c$, for the 
   onset of Taylor--vortex flow in the presence/absence of a magnetic field, 
   and determining the growth/decay rates of either field.
   
   
   An eigenfunction of the linearised equations grows or
   decays exponentially at a rate $\sigma$.   
   \cite{barenghi91} examined convergence with $\dt$ for the
   velocity by comparing against a known growth rate.  
   A simple initial disturbance to the appropriate mode is 
   $\phi \propto x^2(1-x)^2 \sin \alpha z$, 
   or equivalently
   $\phi_{0,\pm 1,0} =  3\Delta$,
   $\phi_{2,\pm 1,0} = -4\Delta$, 
   $\phi_{4,\pm 1,0} =   \Delta$,
   which satisfies the boundary conditions and mimics TVF 
   surprisingly well.
   
   To ensure that the boundary conditions for the magnetic field have 
   been set up correctly we check our method against analytically 
   derived decay rates (see the appendix).  
   Table \ref{tbl:cgce_dt} shows results of the test of growth rates and
   the comparison with \cite{barenghi91}.
   Note that the error is proportional to $\dt^2$.
   \begin{table}
      \bcent
         \begin{tabular}{r@{.}l r@{$\times$}l r@{$\times$}l}
            \multicolumn{2}{c}{$\dt$}	& 
            \multicolumn{2}{c}{\% error in $\sigma$}  &
            \multicolumn{2}{c}{\% error in $\sigma_B$} 	\\[3pt]
            0&01   & \hspace{1.7mm}5.74&$10^{-2}$ & \multicolumn{2}{c}{--}	\\
            0&003  & 5.18&$10^{-3}$ & \hspace{1.7mm} 1.48&$10^{-2}$	\\
            0&001  & 5.76&$10^{-4}$ & 1.65&$10^{-3}$	\\
            0&0003 & 5.2 &$10^{-5}$ & 1.5 &$10^{-4}$	\\
            0&0001 & 6   &$10^{-6}$ & 2   &$10^{-5}$	\\         
         \end{tabular}
         \caption{
            \label{tbl:cgce_dt}
            Error in growth and decay rates. $N\to\infty$.
            For $\eta=1/1.444$, $\alpha=3.13$, $\Rey_1=80$, $\Rey_2=0$ 
            the growth rate is $\sigma=0.430108693$ (Barenghi 1991).
            For the magnetic field $m=1$, $\sigma_B\prandtl=14.055585$,
            $\prandtl=1$.
            The error is proportional to $\dt^2$.
         }
      \ecent
   \end{table}

   To check the interaction of the two fields and the case
   $\prandtl\to 0$ we compare the onset of TVF against \cite{roberts64}.
   Table \ref{tbl:cgce_N} shows the number of modes required to 
   reproduce a few of Roberts' results to five significant figures.
   \begin{table}
      \bcent
         \begin{tabular}{r l r c}
            \multicolumn{1}{c}{$Q$} 	& 
            \multicolumn{1}{c}{$\alpha_c$} & 
            \multicolumn{1}{c}{$N$} 	& 
            \multicolumn{1}{c}{$\Rey_c$} 	\\[3pt]
             30	& 2.69 	&  8	& 280.97	\\
            	& 	& 10	& 281.05	\\
            100	& 1.73	&  8	& 463.20	\\
               	& 	& 10	& 463.52	\\
            300	& 0.928	&  8	& 796.52	\\
            	&  	& 10	& 798.57	\\
                & 	& 12	& 798.55	\\
         \end{tabular}
         \caption{
            \label{tbl:cgce_N}
            Critical Reynolds numbers for varying numbers of modes
            and magnetic field strengths. 
            $\eta=0.95$ with insulating walls.
            For the largest number of modes in each case the values 
            are the same to five significant figures as the results of
            Roberts' (1964) calculations.
         }
      \ecent
   \end{table}

\subsection{\bf Nonlinear two-dimensional flow}      

   The saturation to a steady flow, for some not too large
   $\Rey_1>\Rey_c$, provides a testing ground for the evaluation of 
   nonlinear terms.
   \cite{barenghi91} compares values for velocities at the outflow which
   are in agreement with results obtained by \cite{jones85} using
   a different method.  
   In table \ref{tbl:cgce_NK} we examine convergence with $N$,$K$.
   \begin{table}
      \bcent
         \begin{tabular}{rr r@{.}l r@{.}l r@{.}l}
            $N$	& $K$	&
            \multicolumn{6}{c}{$u_r$} \\[3pt]
            10	& 6	& 4&236577	& 17&94932	& 33&48869	\\
            	& 8	& 4&236615	& 17&97669	& 33&66495	\\
            	& 12	& 4&236616	& 17&97902	& 33&70222	\\
            16	& 6	& 4&233596	& 17&94086	& 33&45839	\\
            	& 8	& 4&233635	& 17&96816	& 33&64135	\\
            	& 12	& 4&233635	& 17&97046	& 33&67982	\\
            -  	& -	& 4&23363	& 17&9705	& 33&6805	\\
         \end{tabular}
         \caption{
            \label{tbl:cgce_NK}
            Radial velocity at the outflow. $\dt\to 0$. 
            $\eta=0.5$, $x=0.5$, $\alpha=3.1631$, 
            and a fixed outer cylinder.
            In the last row are the values for
            $\Rey_1=72.4569,\, 106.066,\, 150.000$
            (Jones 1985).
         }
      \ecent
   \end{table}
   Generally we find that convergence is quicker in $z$ than $r$, and,
   for a given 
   truncation accuracy decreases with increasing $\Rey_1$.  More energy is
   found in the higher modes as $\Rey_1$ is increased.

\subsection{\bf Wavespeeds in wavy TVF}   

   A simple small wavy perturbation
   to axisymmetric TVF that satifies the boundary conditions is
   $\psi \propto x^2(1-x)^2 \sin m_1\theta$ 
   for some wavy mode $m_1$.  
   The purtubation will either decay or grow and saturate depending
   on whether or not the parameters are in the wavy TVF regime.
   \cite{king84} compared wavespeeds founds from physical 
   experiments with numerical calculations.  They found that 
   ``the wavespeed is a sensitive indicator of the accuracy of a 
   numerical code''\dots``any compromise in numerical resolution changes the
   wavespeed by several percent''.  They also argue that the wavespeed can 
   be measured more precisely, both in experiment and numerically, 
   than torques which are dependent on axial wavelength 
   (see the comparison with torque experiments in \S\ref{sect:comp_expt}).
   
   A few of the results used by \cite{marcus84} as a test for his
   numerical method are given in table \ref{tbl:wavespeeds}.
   \begin{table}
      \bcent
         \begin{tabular}{cccc}
            $\Rey_1 / \Rey_c$  & 
            $2\upi / \alpha$  & 
            Marcus  &
            Measured  \\[3pt]
            $3.98$ & $2.40$ & $0.3443\pm 0.0001$ & $0.3440\pm 0.0008$ \\ 
            $3.98$ & $3.00$ & $0.3344\pm 0.0001$ & $0.3347\pm 0.0007$ \\ 
            $5.97$ & $2.20$ & $0.3370\pm 0.0001$ & $0.3370\pm 0.0002$ \\ 
         \end{tabular}
         \caption{
            \label{tbl:wavespeeds}
            Wavespeeds expressed as a fraction of the
            angular velocity of the inner cyliner.
            $\eta=0.868$, $\Rey_c=115.1$, $m_1=6$.
         }
      \ecent
   \end{table}
   Marcus' numerical results were well within the range of 
   experimental error of about 1\%.
   Calculations with our code 
   gave results all within 0.1\% 
   of Marcus' values.

\section{Nonlinear hydromagnetic flow and comparison with experiment}
   \label{sect:comp_expt}

It should first be noted that the results of the previous section
agree well with experiments.  
In this section we directly compare our results with hydromagnetic 
torque experiments.

The torque per unit axial length on the inner cylinder is defined as
\beq
   G = 
   \left.
      \frac{\alpha r}{2\upi}
      \int_0^{2\upi} r \, {\mathrm d}\theta \,
      \int_0^{2\upi/\alpha} {\mathrm d}z
      \left(
         \frac{1}{r} - \pd{r}
      \right) u_\theta
   \right|_{r=R_1} .
\eeq
There is no magnetic torque with insulating boundaries.
For an axisymmetric flow and given the expansion for the velocity
\rf{eq:pot_expansion}, this simplifies to
\beq
   G = 
   \left.
      2\upi r^2
      \left(
         \frac{1}{r} - \pd{r}
      \right)
      \left(
         \psi_0 + \uct
      \right)
   \right|_{r=R_1}
\eeq
The ratio of the effective viscosity of the flow to the kinematic 
viscosity of the fluid is equal to $G/\tilde{G}$ where $\tilde{G}$ is 
the component of the torque due only to the underlying 
circular--Couette flow, $\uct$.

\begin{figure}
   \bcent
      \epsfig{figure=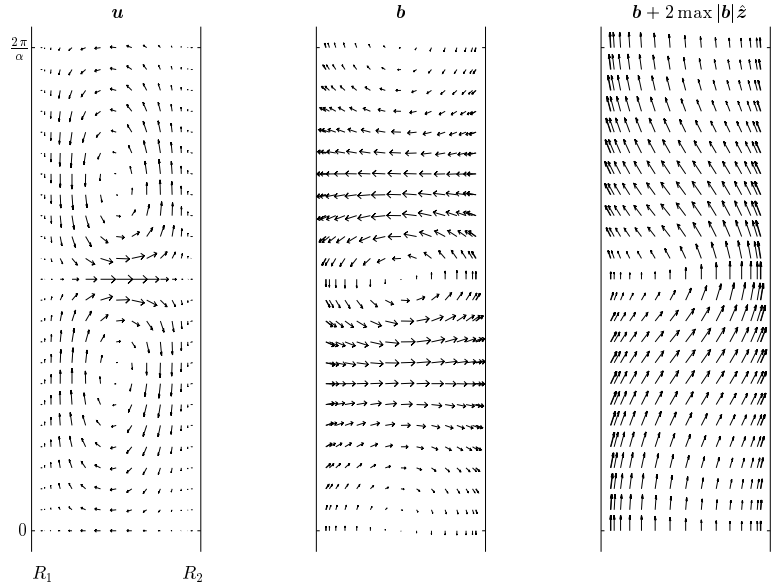, scale=0.71}
      \caption{
         \label{fig:fields}
         Hydromagnetic flow in the presence of an imposed axial field.
         $Q=180$, $\Rey_1=1.5\Rey_c$, $\Rey_c=619$, $\eta=0.95$, 
         $\alpha=1.24$.  The rightmost plot demonstrates that the field
         lines are dragged by the in and outflows.
      }
   \ecent
\end{figure}

Typical nonlinear steady fields in the presence of an imposed axial
fields are shown in figure \ref{fig:fields}.
The main difference between the hydrodynamic and the hydromagnetic
cases is the axial elongation of the Taylor cells.
Not much difference is evident between the eigenfunctions and the
saturated fields other than that the vortex centres move slightly outwards
and closer together towards the outflow.

\begin{figure}
   \bcent
      \epsfig{figure=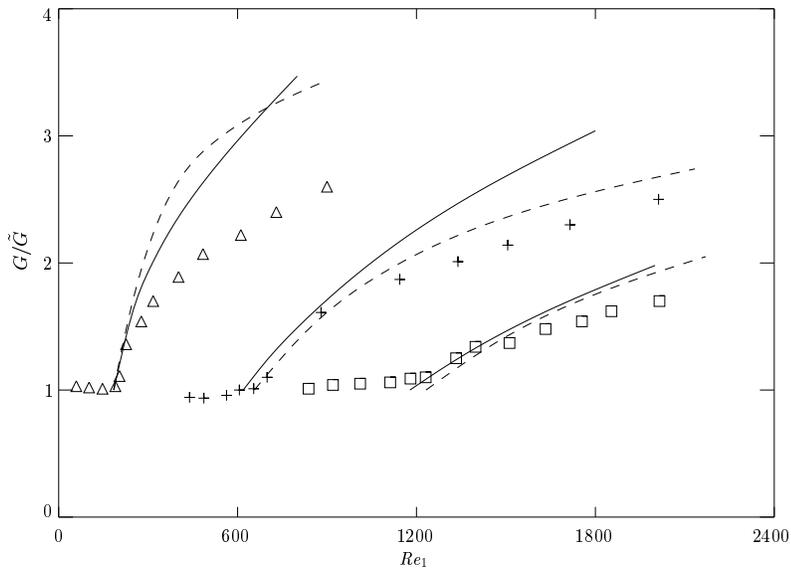, scale=0.71}
      \caption{
         \label{fig:comp_expt}
         Comparison of torques.  
         Experimental results with $\eta=0.95$. 
         $\triangle$, $Q=0$;
         $+$, $Q=180$;
         $\Box$, $Q=652$.
         Solid line, our numerical results.
         Dashed line, Tabeling's expansion about the Reynolds number
         in the narrow gap limit.
      }
   \ecent
\end{figure}

Figure \ref{fig:comp_expt}
shows experimental results obtained by \cite{donnelly62} with mercury and
Perspex cylinders.  Also shown are torques for axisymmetric calculations
with their aspect ratio, $\eta=0.95$, and results
of an amplitude expansion calculated by \cite{tabeling81} 
in the narrow gap limit.  

As the Reynolds number is increased there is good agreement 
between our numerical method, Tabeling's amplitude expansion 
and Donnelly's experiment, until Donnelly's results deviate from 
both ours and Tabeling's calculations.  
The points plotted in figure \ref{fig:comp_expt} are time averages as
significant fluctuations were observed.
Tabeling conjectured that this is due to the appearance of wavy modes. 
With $Q=0$ the onset of wavy modes is not far above the onset of TVF 
and in simulations of these modes we find a reduced torque.

Note that if an axial magnetic field is imposed the onset of wavy modes
is significantly inhibited.  A detailed investigation of the linear
and nonlinear aspects of the wavy modes will be presented in the 
next paper of this series.  Here it suffices to remark the good agreement
between our calculations and the experiment in the weakly nonlinear
axisymmetric regime.

\section{Discussion}

In conclusion we have developed a formulation of
the governing MHD equations of the cylindrical Couette
geometry, suitable for timestepping in the nonlinear regime.  
Results agree well with experiments.

Although the equations do not decouple in the linear part, 
and we must treat mean-flows separately, the formulation 
is similar to that used by \cite{glatzmaier84} in spherical geometry.
We use potentials for the velocity yet do not eliminate the pressure.
This has several advantages.
Our motivation for adopting such a formulation is that the magnetic
field then shares the \emph{same} formulation as the velocity, 
dramatically reducing the potential for error.  
Only a relatively small part of our code is dedicated entirely to the 
magnetic field; this feature is important for testing, as there 
are fewer results against which to compare our results.
Furthermore, it can also accomodate the small Prandtl number limit
with only minor adjustments.
The choice of governing equations which are only second order in $r$
makes the method accurate and matrices easily invertible.  
This feature also enables us to take the same radial
truncation for all variables, if we desire, simplifying implementation 
a great deal.

We have opted to use potentials which ensure divergence-free 
fields.  Primative variable formulations for time integration of the 
Navier--Stokes equations, such as \cite{marcus84} and
\cite{quartapelle95} in this geometry, do not in general extend 
naturally to the magnetic field.  In particular, they are not well suited
to the small Prandtl number limit, relevant to liquid metals available 
in the laboratory.

For the axisymmetric case the expansion by potentials is 
essentially the same as that used by 
\cite{barenghi91} and \cite{jones85}
and results appear to be very similar in terms of accuracy.
Although we have one extra equation, the actual form of our equations
is simpler because we avoided taking the second curl.

Our method is second order in time and exhibits good temporal stability, 
We have not encountered the difficulties experienced by 
\cite{hollerbach00b} and \cite{rudiger00} with 
three-dimensional
potential formulations
and no-slip boundaries.  Using the implict Euler method on the linear
terms reduces the method to $O(\dt/\Rey,\,\dt^2)$.  With 
finite magnetic Prandtl numbers the magneto-rotational instability
(R\"udiger \& Zhang, 2001; Willis \& Barenghi, submitted) 
leads to Reynolds numbers which can be surprisingly low
and the $O(\dt/\Rey)$ error would dominate.

Results obtained using our method compare well with existing 
hydrodynamic literature with respect to the nonlinear equilibration 
of Taylor-vortex flow (Barenghi 1991), the onset of wavy modes (Jones 1985) 
and the wavespeed of wavy modes (Marcus 1984).
In the presence of a magnetic field the results also 
compare well for the linear stability of circular--Couette flow 
(Roberts 1964), and in the nonlinear range the amplitude expansion of
\cite{tabeling81} and experiments of \cite{donnelly62}.

In further work, we will use this method to analyse nonlinear 
three-dimensional hydromagnetic Taylor--Couette flow.
   
\begin{acknowledgments}
   The authors wish to thank Anvar Shukurov and Wolfgang Dobler for 
   stimulating discussions and encouragement during this work, and to 
   the referee for helpful clarifications.
\end{acknowledgments}

\appendix
\section{Decay of the magnetic field}
   \label{sect:mag_decay}

To derive the decay rate of the magnetic field when $\vel=\vec{0}$
we use a mixture of analytical and numerical methods,
different from the numerical technique used to solve the
MHD equations.  We express the magnetic field in terms of two scalar
potentials
   \beq
      \label{eq:mag_tor_pol}
      \magn = \curl(\tor\vechat{z}) + \curl\curl(\pol\vechat{z}) .
   \eeq
Each of $\tor$, $\pol$ are expanded and we seek eigensolutions 
for the magnetic field of the form
   \beq
      A(r,\theta,z) = \sum_{k,m=-\infty}^\infty A_{km}(r) \, 
      \ex^{-\sigma t + \im(\alpha kz + m\theta)},
   \eeq
where $\sigma$ is the decay rate.  Substitution into 
(\ref{eq:non-d_gov}{\it c}) yields the Bessel equation
   \beq
      \label{eq:bessel_eqn}
      \frac{1}{r}\,\pd{r}\,A_{km}(r) + \pd{rr}\,A_{km}(r) +
      \left(
         \hat{\sigma}^2 - \frac{m^2}{r^2}
      \right)
      A_{km}(r) = 0, 
      \quad
      \hat{\sigma}^2 = \sigma\prandtl - \alpha^2 k^2,
   \eeq
   which has solution
   \beq
      A_{km}(r) = 
      A_{km}^J \, J_m(\hat{\sigma}r) + A_{km}^Y \, Y_m(\hat{\sigma}r).
   \eeq
   Matching at the two boundaries the conditions (3.6-8) 
   for the magnetic field defines the problem
   \beq
      \mat{M}(\hat{\sigma}) \, 
      [\tor_{km}^J, \,\tor_{km}^Y, \,\pol_{km}^J, \,\pol_{km}^Y \,]^T
      = \vec{0},
   \eeq
   for the four unknown coefficients.  The quantity
   $\mat{M}$ is a $4\times 4$ matrix and is real if $\hat{\sigma}$ is 
   real (non-oscillatory decay modes).  The slowest decaying 
   eigensolution for $\magn$ is determined by the smallest
   $\hat{\sigma}$ such that $\det \mat{M}(\hat{\sigma}) = 0$.
   Results are shown in table \ref{tab:decay_rates}.
   \begin{table}
      \bcent
         \begin{tabular}{ cc r@{.}l  r@{.}l r@{.}l r@{.}l}
               & 
               $m$ &
               \multicolumn{2}{c}{0} &
               \multicolumn{2}{c}{1} &
               \multicolumn{2}{c}{2} &
               \multicolumn{2}{c}{3} \\
               $k$ & & 
               \multicolumn{8}{c}{$\sigma\prandtl$} \\
            \hline
               0 & $(\tor)$ 
                 & 10&634504 &  2&525434* &  6&259403 & 11&170948 \\
                 & $(\pol)$
                 &  9&613411 & 10&634504  & 13&640533 & 18&474045 \\[3pt]
               1 & $(1^{st})$ & 14&919861 & 14&760834 & 16&619549 \\
                 & $(2^{nd})$ & 20&431404 & 22&010271 & 25&611050 \\[3pt]
               2 & $(1^{st})$ & 45&851458 & 45&868780 \\
                 & $(2^{nd})$ & 49&822104 & 51&678920 \\[3pt]
         \end{tabular}
      \ecent
      \caption{ \label{tab:decay_rates}
         Decay of the magnetic field, $\sigma\prandtl$, 
         for $\eta=0.35$, $\alpha=3.13$.
         When $k=0$ toroidal and poloidal modes separate and their
         slowest decaying modes are given;  $\alpha$ is a redundant
         parameter in this case.  Otherwise they couple and the 
         first two modes are given.  The dominant mode is marked *.
      }
   \end{table}


\end{document}